\documentclass[pra,aps,twocolumn,showpacs]{revtex4}

\usepackage{graphicx,epsfig}
\usepackage{amssymb,amsmath}

\def \g{\gamma}    \def \a{\alpha}    \def \o{\omega}    
\def \b{\beta}    \def \s{\sigma}     
\def \e{\epsilon}          
\def \d{\delta}        \def \l{\lambda}
    \def \o{\omega}

\def \O{\Omega}   \def \S{\Sigma}    
\def \D{\Delta}

\def \h{\hbar}   

\def \f{\frac}
\def \del{\partial}    

\def \ord{\mathcal{O}}

\def\lba{\left(}    \def\rba{\right)}
\def\lbc{\left[}    \def\rbc{\right]}

\def \bra{\langle}   \def \ket{\rangle}

\def\be{\begin{equation}}    \def\ee{\end{equation}}

\DeclareMathOperator{\tr}{tr}

\def \vp{{\bf p}}  
\def \vq{{\bf q}}  
\def \vk{{\bf k}}  

\def \bi{\frac{1}{\beta}}    
\def \nb{f}   
\def \tc{T_{\mathrm c}}   

	\def \c{{\mathrm c}}

\begin{document}

\title{Weakly Non-ideal Bose Gas: Comments on Critical Temperature
Calculations}
\author{Masudul Haque}
\email{masud@physics.rutgers.edu}

\affiliation{Department of Physics and Astronomy, Rutgers University,
136 Frelinghuysen Road, Piscataway, NJ-08854-0849, USA}

\date{\today}

%
%
\begin{abstract}

A number of calculations have appeared for the Bose-Einstein
condensation temperature ($\tc$) of a weakly repulsive dilute Bose
gas.  After a short survey of previous work, I point out several
issues related to these approaches, and outline future calculations.

\end{abstract}

\maketitle

\section{Introduction}   \label{sect_intro}

During the recent upsurge of activity in the field of Bose-Einstein
condensation (BEC), some older questions concerning the weakly interacting
Bose gas have been revived.  One such un-resolved question is the
effect of a weak repulsion on the critical temperature $\tc$.
Calculations on this question has produced widely dissimilar results:
increases and decreases of $\tc$ proportional to $a$, $\sqrt{a}$,
$a\ln{a}$, etc. have been reported by various authors.  (Here $a$ is
the scattering length.)  It is only during the past 2-3 years that the
community has been converging toward a consensus.

Considering the diverse approaches to the $\tc$ problem, and the
variety of issues that have arisen in this connection, it seems
appropriate at this point to attempt to draw some connections between
the different approaches, and point out some misconceptions and fine
points.  The purpose in this article is to review and comment on
existing work, rather than to present new calculations.

The model is described by the Hamiltonian
\[
\hat{H} = \sum_{\vk} \e_{\vk} \hat{b}_{\vk}^\dag \hat{b}_{\vk}
+  \frac{1}{2V}\sum_{\vp,\vq,\vk} U(\vk)\, \hat{b}_{\vp+\vk}^\dag
\hat{b}_{\vq-\vk}^\dag \hat{b}_{\vp} \hat{b}_{\vq} \, \, ,
\] 
where $\e_{\vk}=k^2/2m$ is the free-gas spectrum, and $\hat{b}$,
$\hat{b}^\dag$ are bosonic operators.  In principle, one is interested
in any possible interaction function $U(\vk)$, but the simplest
possible form is a delta function in real space, so that $U(\vk) = U$
is momentum-independent.  To first order $U$ is related to the
$s$-wave scattering length $a$ by $U = 4\pi\h^2a/m$.  A dimensionless
measure of the interaction is the quantity $an^{1/3}$, where $n = N/V$
is the density.  The program is to examine the shift in the critical
temperature $\tc$, as compared to a noninteracting gas of the same
density, as a function of $an^{1/3}$, for small $an^{1/3}$.

In section \ref{sect_background}, the background is set by summarizing
different calculations and basic results.  Section \ref{sect_heart}
contains the comments on various approaches to the $\tc$ problem.  In
the concluding section \ref{sect_further}, I suggest some directions
for additional calculations.

\section{Background and Perspective}  \label{sect_background}

Early predictions for the shift in transition temperature appear in
the work of Lee, Yang and Huang, around forty years ago, as summarized
in \cite{baym_bigpaper}.  Recently, Huang has published
\cite{huang_prl} a prediction for an increase in $\tc$ due to the
inter-particle repulsion, proportional to the square root of the
interaction strength, $\tc-\tc^{(0)} \sim \sqrt{an^{1/3}}$.  Schakel
in a short note \cite{schakel_comment}, and Baym {\it et al} in
\cite{baym_bigpaper}, have both discussed the problem with this work;
in particular \cite{baym_bigpaper} contains an analysis showing how
Huang's result is generated spuriously from a truncation of the virial
expansion used at mean field level.

A similar spurious result appearing in the literature is that of
Toyoda \cite{toyoda}, who in 1982 performed mean-field-like
calculations deriving a Landau-Ginzburg type thermodynamic function.
His prediction was a \emph{decrease} of the transition temperature,
$\D\tc \sim - \sqrt{an^{1/3}}$.  Again, this is believed to be an
artifact of mean field approximations.  The present understanding is
that there is no mean field effect on the transition temperature, and
that higher-order correlations are required to find the leading shift
in $\tc$.

A more recent work that also predicts a negative shift in $\tc$ is the
canonical-ensemble work of Wilkens \emph{et al} \cite{wilkens}.  This
work has been critiqued in \cite{mueller_baym_finite}. I discuss some
related issues in section \ref{sect_canonical}.

In addition, in view of the continuing appearance of negative-shift
predictions, I give in section \ref{sect_direction+IR} a simple
argument for the \emph{increase} of the transition temperature due to
the addition of a repulsive interaction.  This argument makes use of
eq \eqref{formula}, which is derived independent of any particular
method of treating interaction effects.

During the past several years, a large body of work has been published
by Baym, Lalo\"e and collaborators
\cite{baym_prl,baym_N,baym_a2log,HGL_ursell,mueller_baym_finite,baym_bigpaper,baymfriends_numerics}, and a
number of authors have done followup work
\cite{arnold_N,arnoldmoore,svistunov}.  These workers predict a
positive shift in $\tc$ that to leading order is \emph{linear} in the
scattering length; $\D\tc = {c_1}an^{1/3} +
\ord(a^2n^{2/3})$.  Attempts to calculate the coefficient
$c_1$, however, have continued to give fluctuating results, and this
has prompted the Baym group to predict \cite{baym_a2log} a logarithmic
contribution at the \emph{next} order, i.e., a $-a^2\ln{a}$
contribution.  The presence of such a non-analyticity might explain the
difficulty in any numerical estimate of $c_1$.

Stoof's renormalization-group analysis \cite{stoof1,stoof2} of the
three-dimensional bose gas has led to a prediction
\cite{stoof2,stoof_private} of an increase of the transition
temperature proportional to $a|\ln{a}|$, i.e., a non-analyticity at
\emph{leading} order.  A leading non-analytic behavior would actually
explain even better the uncertain results obtained in attempts to
calculate the coefficient $c_1$, based on the assumption of linear
shift $\D\tc \sim {c_1}an^{1/3}$.

An older attack on the $\tc$ problem is that of Kanno
\cite{kanno12,kanno3}.  In a series of papers, Kanno introduced a
``quasilinear'' canonical transformation approach to the many-body
problem, which he used to calculate (among other things) the free
energy and hence transition temperature of the bose gas.  Since
Kanno's work has mostly been ignored by the current $\tc$ community, I
give a description of his work in section \ref{sect_kanno}.  I also
show how some of his derivations can be performed in a more
transparent way.

Schakel's work on developing an effective theory \cite{schakel_ijmp}
of the weakly non-ideal Bose gas has led to the prediction
\cite{schakel_boulevard} of a shift in the transition temperature that
is strictly linear in the interaction.  The connection of this calculation
with other approaches has not yet been explored in detail.  Some
issues are pointed out in section \ref{sect_schakel}.

Our recent report \cite{our_f02} addresses the question of $\tc$
within a quasiparticle framework.  Other than working out details of a
quasiparticle description (ultraviolet divergence, effect of
quasiparticle lifetime), we also provide a non-selfconsistent method
of regularizing the infrared divergences appearing in the description
of the transition point.  This results in a shift of $\tc$
approximately $\propto$~$a\sqrt{|\log{a}|}$.  The procedure is
uncontrolled; however, this result points strongly to a non-analytic
contribution in the dependence of $\tc$ on the interaction.

Other relevant calcultaions include Kleinert's ``five-loop''
determination \cite{kleinert} of the coefficient $c_1$ mentioned
above, assuming a puely linear leading shift, and the ``linear
$\d$-expansion'' calculations \cite{delta} of de Souza Cruz \emph{et
al}.  Kleinert has questioned \cite{kleinert} the $\d$-expansion
method, but a detailed analysis of the relationship of this approach
to others seems to be lacking in the literature.

\subsection{Mean Field}   \label{sect_meanfield}

For a momentum-independent potential, the two mean field contributions
to the self-energy (Hartree and Fock) are equal, $\S^{(1)}(\vk,\o) =
Un + Un = 2Un$.  Since the self-energy is momentum-independent, it can
simply be absorbed into a redefinition of the chemical potential.
\begin{gather} 
E_{\vk}^{(1)} = \frac{k^2}{2m} - \mu + \S^{(1)}(\vk,E_{\vk}^{(1)}) =
\frac{k^2}{2m} - \mu'   \\ 
N = N_0 + \frac{V}{\l^3}g_{3/2}(e^{\b\mu'})   \nonumber 
\end{gather}
Here $\mu' = \mu-\S^{(1)} = \mu-2Un$.
The transition now simply corresponds to $\mu' =0$ instead of $\mu
=0$, and $\tc$  remains unchanged.  The lesson here is that a
momentum-dependent self-energy is required for a shift in $\tc$.

In second-order calculations, (e.g. in \cite{our_f02},) it is usual to
have the 1st-order self-energy already absorbed in the chemical
potential, i.e., to use $\mu'$ as the chemical potential.  This allows
one to exclude those second-order diagrams which can be obtained by
1st-order self-energy insertions into 1st-order diagrams.

\emph{Spatially extended interaction}.
In a real system, of course, the description of the interaction in
terms of a single parameter ($U_0$ or $a$) is valid only for very weak
interactions or at very low temperatures/densities; if these
conditions are not satisfied one has to take into account the detailed
shape of the potential $V(x) = V_0\s(x)$ which is not a delta function
in real space (not a constant in momentum space) any more.  In such a
case we get a momentum-dependent Fock term, and hence a contribution
to $\tc$ from the first-order self-energy.  Assuming a Gaussian $V(x)$
of width $a$ and height $V_0$, one gets a correction to the spectrum
that can be incorporated as an effective mass $m^* =
m[1-V_0a^2nm/3]^{-1}$, resulting in a negative shift \cite{fetter}:
$$
\f{\D\tc}{\tc^{(0)}} = - \f{1}{3}mn V_0a^2 \, \, .
$$
The negative shift at higher densities is present in liquid helium and
presumably in other Bose fluids.  However, for weak interactions, this
shift is less important than the leading $\ord(a)$ effect, which is
the current theoretical challenge.

\subsection{$\tc$ in Liquid Helium}

As a highly correlated liquid, the properties of liquid helium are not
expected to be described well by the dilute weakly-interacting Bose
gas model.  However, the deviation of the lambda transition
temperature (2.2K), from the BEC temperature of an ideal gas with the
same density (3.1K), can be mostly explained using a simple effective
mass description.  In other words, the experimental $\tc$ is
reproduced quite closely by replacing the helium atomic mass $m$ by a
quasiparticle effective mass $m$* in the ideal gas expression
$\tc^{(0)} = \lba{2}\pi/m[\zeta(\tfrac{3}{2})]^{2/3}\rba n^{2/3}$.

The effective mass in the liquid exceeds the bare mass, because of the
inertia of the medium which has to make way for any atom to move.
This results in a lowering of $\tc$.  The effect is essentially
equivalent to that described in the last section for a
momentum-dependent interaction.

Feynman \cite{feynman53} calculated a partition function for liquid
helium that takes into account the effective mass enhancement
described above.  Using this partition function, Chapline calculated
\cite{chapline} a $\tc$ quite close to the actual $\l$-point of
helium.  This is equivalent to using the effective mass in the ideal
gas expression for $\tc^{(0)}$.

According to current understanding, therefore, one expects the
following variation of $\tc$ with $n^{1/3}a$: an initial linear-like
increase, followed by a maximum and then a decrease due to the
effective-mass mechanism described above, until eventually $\D\tc$
becomes negative.  This kind of dependence has been seen in the
Monte-Carlo calculation of Gr\"uter, Ceperley, Lalo\"e
\cite{baymfriends_numerics}, and in Reppy's experiments \cite{reppy}
described below.

\subsection{Relevant Experiments}

One could think of studying the $\tc$ problem experimentally, e.g., in
the context of trapped atomic Bose gases.  Unfortunately, trap effects
dominate in the dependence of $\tc$ on the interaction for these
systems.  Turning on a repulsion for a trapped Bose gas causes an
expansion of the gas, reducing the density and hence decreasing the
transition temperature $\tc\propto{n}^{2/3}$ \cite{trap_tc}.  In
contrast to the intrinsic effect under review here, trapping effects
on $\tc$ are found at mean-field level already.

A different experimental realization of the weakly non-ideal bose gas
is the $^4$He-Vycor system explored by Reppy \emph{et al}
\cite{reppy}.  Here the small-$an^{1/3}$ regime is reached by
enormously decreasing the $^4$He density.  Reppy's investigation of
the transition temperature shows a linear-like increase of $\tc$ at
low $an^{1/3}$, followed by a maximum and a decrease to values lower
than $\tc^{(0)}$.  The general behavior is satisfying, but for the
initial increase, the data is not yet good enough to choose between
$\sim{a}$, $\sim{a}\ln{a}$, or $\sim{a}\sqrt{\ln{a}}$-like behavior.

\section{Comments on Various Approaches}  \label{sect_heart}

\subsection{Baym and collaborators}

The work of Baym, Lalo\"e and collaborators
\cite{baym_prl,baym_N,baym_a2log,
mueller_baym_finite,baym_bigpaper,HGL_ursell,baymfriends_numerics} 
establishes definitely that a repulsive interaction \emph{increases}
$\tc$, and that the dependence of $\D\tc$ on the interaction is of
linear order, as opposed to quadratic or $\sim\sqrt{a}$.  

One should note, however, that the Baym group's work relies heavily on
power-counting and dimensional arguments.  This kind of argument
typically does not catch logarithmic corrections.  The possibility of
leading non-analytic dependence of $\D\tc$ on the interaction,
therefore, needs to be further examined.

It has been suggested \cite{schakel_private} that the large-$N$
calculation \cite{baym_N} is a reliable indicator that the leading
shift is purely linear, because of intuition gained from other systems
where large-$N$ calculations capture all logarithms.  I would like to
point out, however, that the $N\neq{2}$ models belong to a different
universality class, and have different excitation spectra as compared
to the usual ($N=2$) Bose gas.  More care is therefore required in
interpreting results from the large-$N$ approach.

The Baym-Lalo\"e collaboration seems to have been the first to relate
the single-particle spectrum at $\tc$ to the shift $\D\tc$.  Another
noteworthy aspect is the use of Ursell operators
\cite{HGL_ursell,baym_bigpaper}, and a corresponding diagram
technique, to describe particle correlations.  (As far as I can tell,
for this problem, the Ursell operator technique does not give any
extra information compared to the usual Green function techniques.)

\subsection{Direction of Shift, Infrared Spectrum}
\label{sect_direction+IR} 

Once the shift in $\tc$ is related to the modification of the
single-particle spectrum, the direction of the shift can be argued
using well-known facts about the spectrum of the Bose gas.

We will use the following expression for the shift, as appearing in
\cite{our_f02}:
\begin{equation}    \label{formula}
\frac{\D\tc}{\tc^{(0)}} ~\approx~
-\frac{2}{3n}  \int \frac{d^3  k }{(2\pi)^3} \ 
\lbc  f_\c (\xi_\vk) ~-~ f_\c (\e_\vk)  \rbc  \, \, .
\end{equation}
Here $f_\c$ is the Bose function at the critical point; $f_\c(x) =
(e^{x/k\tc}-1)^{-1}$; and $\xi_\vk$ is the quasiparticle spectrum
at $\tc$, i.e., $\xi_\vk = \e_\vk + \S(\vk,\xi_\vk;T=\tc)$.  

The expression used by Baym \emph{et al}
\cite{baym_prl,baym_bigpaper}, $\D\tc \propto
\int{dk}U(k)/[k^2+U(k)]$, is the high-temperature (zero Matsubara
frequency) version of \eqref{formula}.  Noting that the major
contribution to \eqref{formula} comes from the infrared, one can write
approximately
\begin{equation}  \label{formula_highT}
\D\tc \propto \int_0^{\rm cutoff} dk\ k^2 [\e_\vk^{-1}-\xi_\vk^{-1}] 
\, \, .  
\end{equation}
This equation might introduce or leave out logarithmic corrections;
however the following argument is unchanged for \eqref{formula}.

If the quasiparticle spectrum $\xi_\vk$ is ``harder'' than the bare
spectrum $\e_\vk = k^2/2m$, i.e., if $\xi_\vk$ is sub-quadratic in $k$
or quadratic with an effective mass $m^*<m$, then the $\e_\vk$
integral dominates and the shift is positive.  On the other hand if
the spectrum at the transition point is ``softened'' by the
interaction, then eq \eqref{formula_highT} [ or \eqref{formula}] leads
to a decrease in the transition temperature.

Since the weakly interacting bose gas is known to have a linear
(Bogoliubov) infrared spectrum below the critical temperature, and a
quadratic spectrum above $\tc$, it is extremely reasonable to assume,
even without any RG or perturbation-theoretic calculations, that the
spectrum at the critical point should be something intermediate
between linear and quadratic.  As pointed out above, this means a
\emph{positive} shift in the transition temperature.

The long-wavelength spectrum at the transition point should be
$\xi_\vk \sim{k}^{2-\eta}$, where $\eta$ is the anomalous dimension.
The present system falls in the same universality class as the $XY$
model, or the $N=2$ quantum rotor model.  For this universality class,
large-$N$ and $\e$-expansion calculations give respectively
$\eta\approx{0.14}$ and $\eta\approx{0.02}$, and a recent Monte Carlo
calculation \cite{eta_montecarlo} has produced the value
$\eta\approx{0.038}$.  

Second-order treatments of the self-energy (e.g., the Baym group's
\cite{baym_prl,baym_bigpaper} and our \cite{our_f02} work)
over-estimates the modification of the spectrum, giving a $k^{3/2}$
infrared behavior ($\eta=0.5$). On the other hand, some other
treatments (such as Stoof's \cite{stoof2} and Schakel's
\cite{schakel_ijmp,schakel_boulevard}) neglect the modification
altogether ($\eta=0$).

\subsection{Renormalization group treatment (Stoof)}

The RG flow equations derived by Bijlsma \& Stoof \cite{stoof2} allow
one to numerically locate the transition point.  Under RG flow, the
physical effective chemical potential flows to positive and negative
values, if one starts out respectively in the Bose-condensed and the
symmetry-unbroken phase.  The critical point is therefore identified
by finding parameters for which the chemical potential flows to zero.
Stoof finds \cite{stoof2,stoof_private} that the dependence of $\D\tc$
on $a/\l$ has more structure than purely linear; $\D\tc \propto
(a/\l)\ln(a/\l)$.

The differences between Stoof's and the Baym group's results merit
further study.  Stoof has suggested \cite{stoof_private} that the
source of the difference is not the infrared but the
\emph{ultraviolet}.  The correction to the spectrum, $U(k) \propto
\S(\vk,E_\vk) - \S(0,0)$, has an ultraviolet $\ln{k}$ behavior in the
Baym group's treatment \cite{baym_bigpaper,baym_prl}.  In the RG
framework, the corresponding quantity is the running coupling constant
$\mu(\Lambda)$, which shows no ultraviolet logarithms in the
calcultation of ref \cite{stoof2}.  This leads to a logarithmic
difference in the two predictions for $\tc$.  Stoof also suggests
\cite{stoof_private} that an improved RG calculation, including the
momentum dependence of the effective two-body interaction at the
transition point, would shed light on the difference between the two
approaches.

\subsection{Schakel: IR regularization}  \label{sect_schakel}

In treating his effective-action theory near the transition point,
Schakel \cite{schakel_ijmp,schakel_boulevard} is also faced with
infrared (IR) divergences.  One of the remarkable features of this
work is an innovative analytic continuation procedure used to regulate
IR problems.

Baym {\it et al} have remarked \cite{baym_bigpaper} that Schakel's
prescription ($\zeta$-function regularization) must be incorrect
because it gives an incorrect result for the compressibility of the
ideal gas.  This argument, unfortunately, is derived from an equation
that Schakel derived explicitly for the nontrivial broken-symmetry
ground state \cite{schakel_boulevard,schakel_private} of an
interacting gas ($\mu>{0}$).  The same equations are not applicable to
the noninteracting case ($\mu\leq{0}$) for which the structure of the
vacuum is different (more trivial).  The situation is therefore more
complicated than indicated in Baym {\it et al}'s note.  In response,
Schakel has provided a more detailed explanation and justification
\cite{schakel_rejoinder} for the $\zeta$-function continuation
procedure.

It seems to me that the regularization procedure itself is quite
possibly sound.  My concern with Schakel's derivation is different ---
just below the transition point (broken-symmetry phase), Schakel uses
a quadratic spectrum.  A more accurate treatment would use the
phonon-like spectrum $E_\vk \approx \sqrt{\e_\vk^2+2Un_0(T)\e_\vk}$.
The effect such a spectrum would have on the thermodynamic potential
or on $\tc$ is not clear.  (Redoing Schakel's $\tc$ calculation with
a phonon-like spectrum seems quite difficult.)

\subsection{Canonical Ensemble: Role of $N_0(\tc)$.}
\label{sect_canonical}

In Wilkens {\it et al}'s paper \cite{wilkens}, calculations have been
performed using a canonical ensemble of $N$ particles, and the shift
$\D\tc$ obtained is linear and \emph{negative} in the scattering
length.  It has been pointed out \cite{mueller_baym_finite} that the
error lies in evaluating quantities (derivatives) at $N_0 = 0$,
instead of evaluating them at the expectation value of $N_0(\tc)$.
Mueller {\it et al} \cite{mueller_baym_finite} have also used
finite-size scaling ideas in combination with a first-order
calculation in the canonical ensemble, and obtain a positive linear
shift.

The idea that the zero-momentum state occupancy $N_0$ becomes
``microscopic'' at and above the transition needs to be clarified.
One finds $\lim_{N\to{0}} (N_0/N) = 0$ for $T\geq \tc$, where $N$ is
the total number of particles, but not necessarily $N_0 \sim \ord(1)$.
In fact, for both the ideal and the interacting Bose gas, $N_0(\tc)
\sim N^{2/3}$, which is large compared to $1$ though small compared to
$N$.

In view of the continuing appearance of the $N_0(\tc) \sim \ord(1)$
misconception, I give here a simple derivation of $N_0(\tc) =
{\g}N^{2/3}$, together with an expression for the coefficient $\g$,
for the ideal-gas case.  

Let us consider the ideal-gas relationship $N = N_0 +
\frac{V}{\l^3}g_{3/2}(e^{-\a})$, where $\a=-\b\mu$, and $g_\nu$ refers
to a Bose-Einstein integral function \cite{BE_functions}.  Since $\a$
is small near the transition, one can use the Robinson expansion
\cite{BE_functions} for $g_{3/2}$.  Also, inverting $N_0 =
1/(e^{\b\e_0+\a}-1)$ one gets $\a \approx 1/N_0$.  Combining these
yields, near the transition,
\[
N_0^{3/2} 
+ \lbc\frac{V}{\l^3}\ \zeta \lba\tfrac{3}{2}\rba -N \rbc \sqrt{N_0}
- \frac{V}{\l^3} (2\sqrt{\pi})  ~\approx~ 0 \, \, .
\]
Since $n\l^3 = \zeta(\tfrac{3}{2})$ at the transition, the coefficient
of $\sqrt{N_0}$ needs to vanish.  Therefore
\begin{equation}  \label{N0_at_tc_ideal}
N_0(\tc)  ~\approx~ 
\lbc \frac{2\sqrt{\pi}}{\zeta(3/2)} \rbc ^{2/3}   N^{2/3}  \, \, .
\end{equation}
The $\vk=0$ occupancy $N_0$ can actually be calculated for all
$T$, using an extra Lagrange multiplier $\mu_0$ conjugate to $N_0$, in
addition to usual chemical potential $\mu$ coupling to $N$. Using this
``two chemical potentials'' formalism \cite{our_unpubl}, one can show
that for the ideal gas $\langle{N_0}\rangle =1$ occurs at a
temperature $T\approx {2.6}\,\tc^{(0)}$, well above the transition
point. 

For the interacting-gas case, an equation corresponding to
\eqref{N0_at_tc_ideal} is difficult to derive, but finite-size scaling
arguments \cite{cardy_book,mueller_baym_finite} show that $N_0\sim
N^{2/3}$ continues to hold, even though the nonideal Bose gas belongs
to a different universality class.

The definition of the transition point used in ref \cite{wilkens} is
in terms of the probability
\begin{equation}  \label{canonical_Pn}
P_n(T,N) \propto \tr 
\lbc \d_{\hat{N}_0,n} e^{-{\b}\hat{H}} \d_{\hat{N},N} \rbc 
\end{equation}
for having a condensate occupancy $N_0 =n$.  The claim is that
$P_n(T,N)$ decreases monotonically with $n$ for all temperatures above
the transition, and has a peak at nonzero $n$ for in the condensed
phase, and therefore the transition should be defined as the
thermodynamic limit of the zero of $D(T,N) = P_0(T,N) - P_1(T,N)$.  

The correspondence between the canonical particle-counting statistics,
and the grand-canonical ideas of average occupancies, is not trivial.
One important point to note is that a monotonically decreasing $P_n$
does not imply a zero expectation value for $N_0$, even though $P_n$
has a maximum at $n=0$.  (For example, using $P_n\propto e^{-{\a}n}$
gives $\bra{N_0}\ket = \sum_n nP_n = 1/[e^\a-1]$.)  The stipulation of
a monotonically decreasing $P_n$ therefore does not imply a zero or
$\ord(1)$ value of $N_0$ at the transition.

\subsection{Kanno's Approach: Using Modified Thermodynamic Potential}
\label{sect_kanno}

Kanno developed an unusual approach \cite{kanno12} to study many-body
systems, and also used this approach to calculate Bose gas properties
\cite{kanno3}.

\emph{The formalism}.\\
Kanno's calculation of thermodynamic quantities involves a
``quasilinear'' canonical transformation \cite{kanno12} of the
free-particle operators $\hat{b}$ ($\hat{b}^\dag$):
\begin{multline}  \label{kanno_quasilinear}
\hat{b}_i = \hat{\g}_i 
+ \sum_{j;k,l} A_i(j;k,l) \hat{\g}^\dag_j \hat{\g}_k \hat{\g}_l  \\
+ \sum_{j,k;l,m,n} A_i(j,k;l,m,n) \hat{\g}^\dag_j \hat{\g}^\dag_k 
\hat{\g}_l \hat{\g}_m \hat{\g}_n   + \ldots  \, \, ,
\end{multline}
resulting in a Hamiltonian having a particular (diagonal-like) form.
The formalism involves examining the properties of the coefficients
$A$, and the coefficients appearing in the quasi-diagonal hamiltonian.
Applying this formalism to the weakly repulsive bose gas
\cite{kanno3}, one can derive the Bogoliubov spectrum, entropy and
pressure above and below the transition, and many other quantities.

Kanno does not provide direct connections to the usual language of
perturbative many-body theory.  I would like to point out that many of
his physical results, if not all, are equivalent to results obtained
in more mainstream approaches.  For example, the quantities $A(\vp)$
and $B(\vp)$, appearing in the second paper of the series
\cite{kanno12}, are equivalent to the normal and anomalous
self-energies that appear in modern descriptions.

\emph{The shift in $\tc$}.\\
Kanno's defines \cite{kanno3} a modified thermodynamic potential
\[
\O'(N_0, \mu', T) = - \bi 
\ln \tr' e^{-\b(\hat{H}-\mu'\hat{N'})} \, \, ,
\]
with an unusual Lagrange multiplier $\mu'$ conjugate to the number of
non-condensate particles, $N' = N-N_0$, as opposed to the usual
chemical potential which is conjugate to the total particle number
$N$.  It follows that
\begin{equation}  \label{kanno_N}
N - N_0 ~=~ N' ~=~ - \frac{\del\O'(N_0,\mu',T)}{\del\mu'} \, \, .
\end{equation}
Once $\O'$ is calculated in some approximation, comparing this
relation to $N-N_0 = Vg_{3/2}(e^{\b\mu})/\l^3$ for the free bose gas
at the transition point yields the shift $\D\tc$, in terms of the
quantity $\mu'(\tc)$.  

The quantity $\mu'$ itself is obtained from a consideration of the
free energy $F(N_0,N,T) = \O'(N_0,\mu',T) + \mu'N'$.  The actual value
of $N_0$ chosen by the system is the one that minimizes $F(N_0,N,T)$,
i.e.,
\begin{equation}   \label{kanno_mu'}
0 = \left. \frac{\del F}{\del N_0} \right|_{T=\tc} 
=  \left. \frac{\del \O'}{\del N_0} \right|_{T=\tc}  -\mu'(\tc) \, \, .
\end{equation}
This equation allows one to express $\mu'(\tc)$ in terms of the
interaction strength.

\emph{Computing $\O'$ in more usual language}.\\
As advertised in section \ref{sect_background}, I show here how the
modified thermodynamic potential $\O'$ (which is the quantity relevant
to the $\tc$ problem) can be obtained from a diagrammatic second-order
perturbative calculation.

The Feynman rules for a diagrammatic perturbative calculation of the
modified therodynamic potential $\O'(\mu',N_0,T)$ are very similar to
those for the usual $\O(\mu,T)$.  One has to use a momentum-dependent
chemical potential $\mu_\vk = \mu'(1-\d_{\vk,0})$, and separate out
$\vk=0$ from the momentum-sums at each stage.  Two first order and
two second-order diagrams need to be calculated:
\begin{widetext}
\[
\O' \, \approx \, \O'^{(0)} + 
\begin{minipage}{2cm} \begin{center}
\epsfig{file=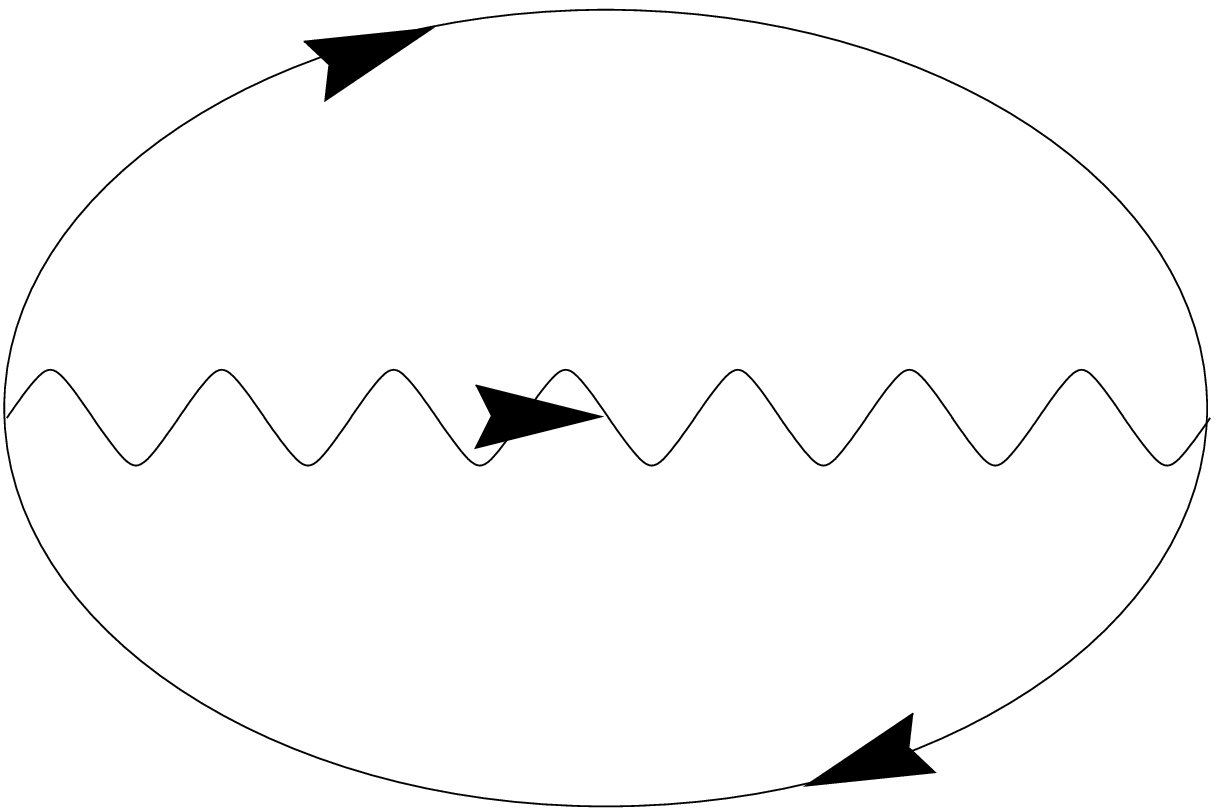,width=1.5cm}\end{center}\end{minipage} + 
\begin{minipage}{2cm}  \begin{center}
\epsfig{file=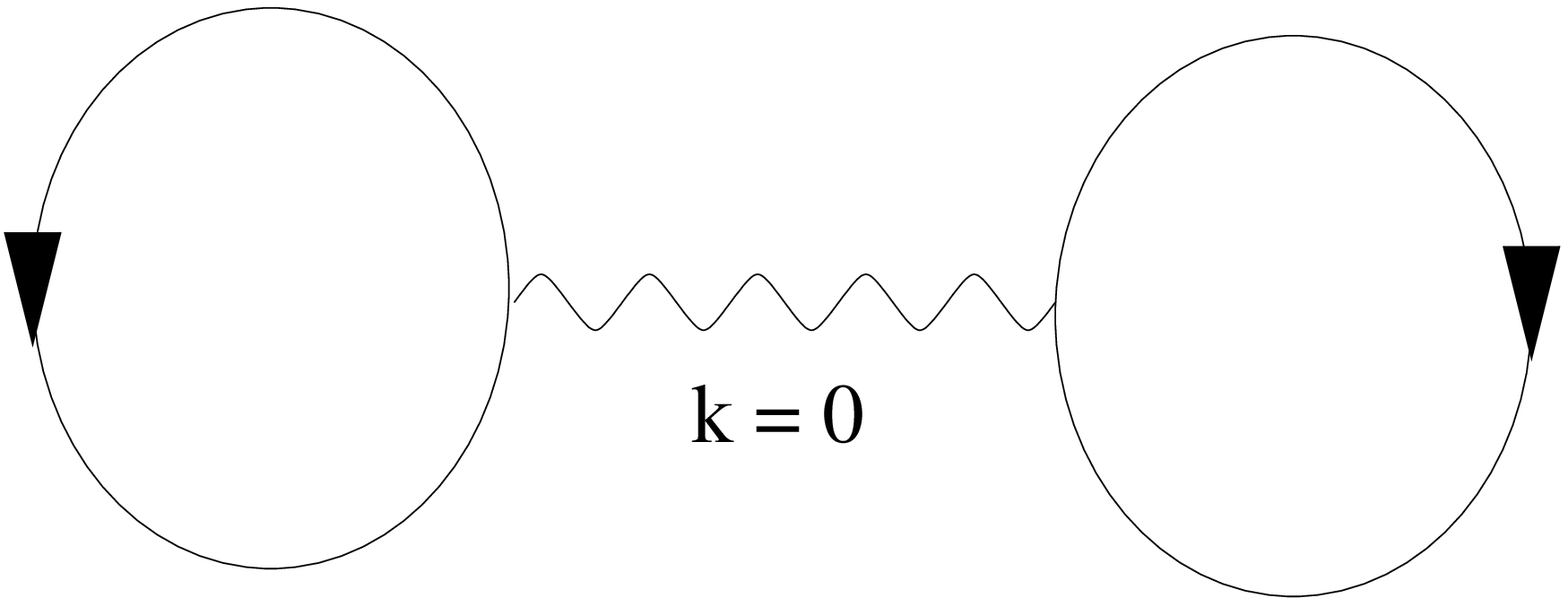,width=1.5cm, height=0.8cm} 
\end{center}\end{minipage}  +
 \begin{minipage}{2cm}  \begin{center}
\epsfig{file=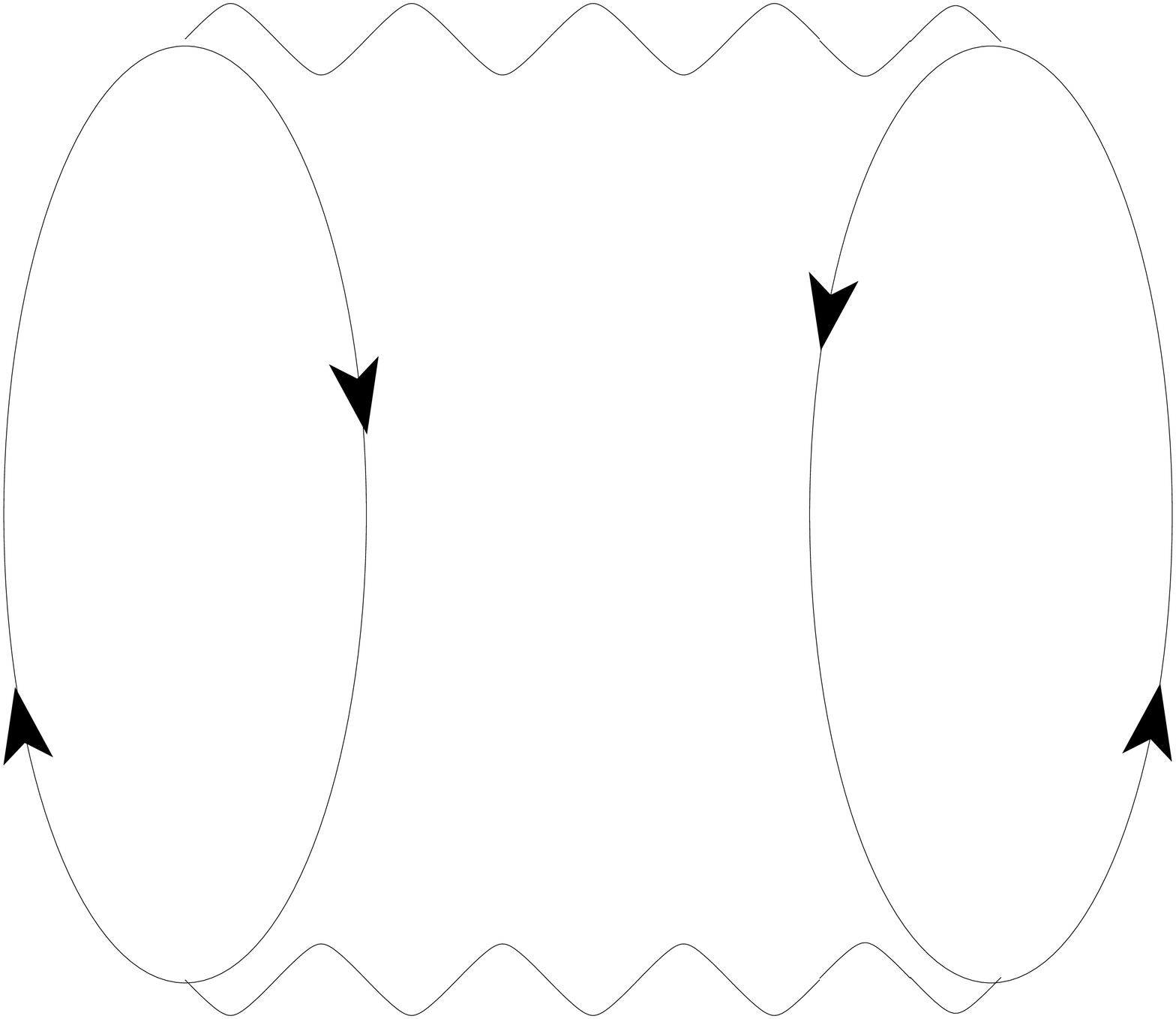,width=1.5cm} \end{center}\end{minipage}   +
\begin{minipage}{2cm}  \begin{center}
\epsfig{file=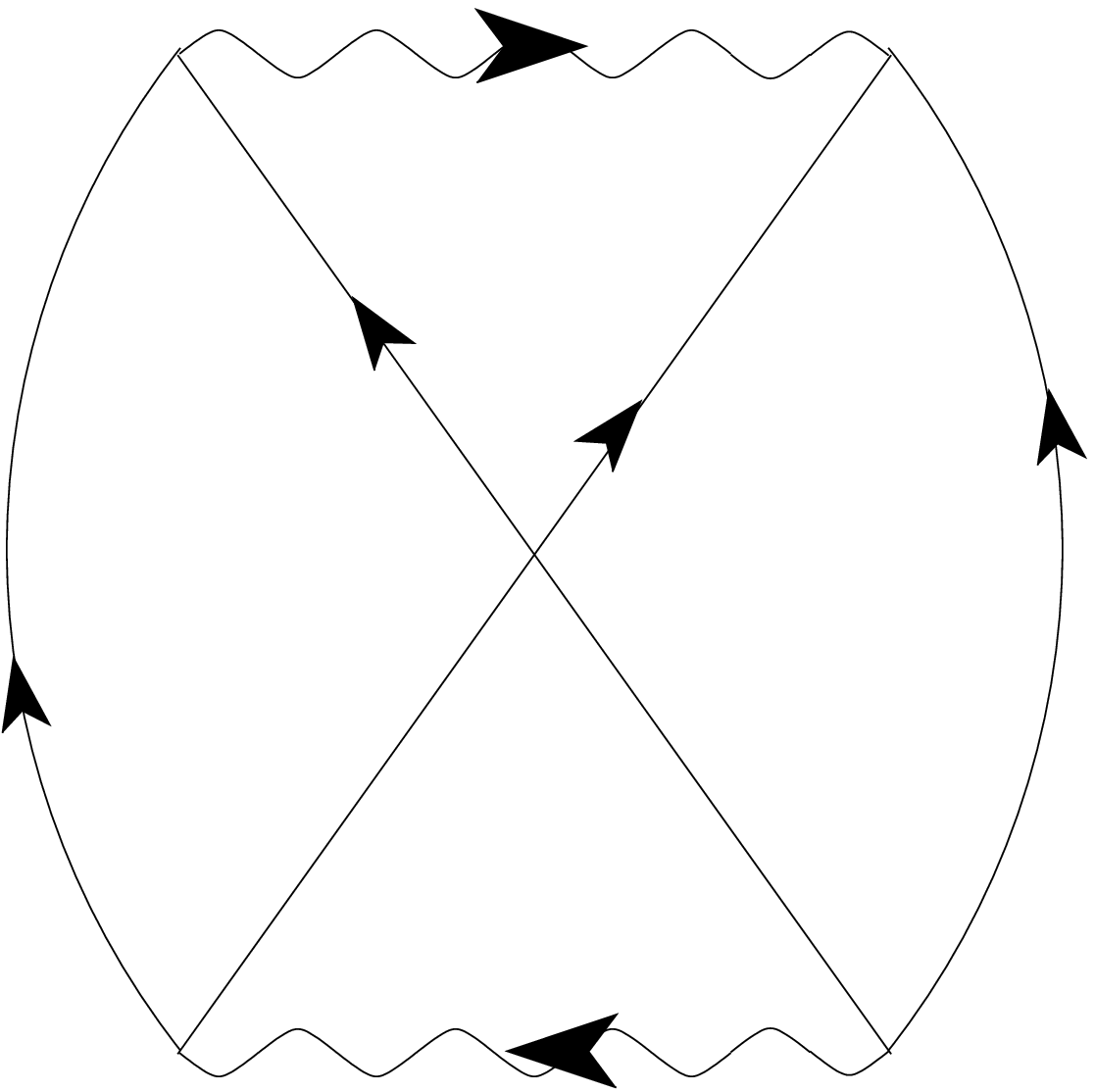,width=1.5cm,height=1.4cm}
\end{center}\end{minipage}  
\, \, .
\]
Other second-order diagrams are excluded by considering the chemical
potential to be renormalized upto first order, as discussed in section
\ref{sect_meanfield}.

The zeroth and first order terms can be evaluated exactly.  UV
divergences appearing at second order can be removed by eliminating
the bare potential $U$ in favor of the 2-body $t$-marix, as done for
the self-energy in \cite{our_f02}.  The result is
\begin{multline}  \label{kanno_O'_prtbn}
\O' =   - \frac{V}{\b\l^3}g_{5/2}(e^{\b\mu'})  + \frac{t}{V}N^2 
+ 4 t^2 N_0 \int_{\vp} \int_{\vq} \frac{\nb(\e_{\vp} -\mu')  
\nb(\e_{\vq} -\mu')}{\e_{\vp} -\e_{\vq} -\e_{\vp -\vq}}   \\
  +  2 t^2 V \int_{\vp} \int_{\vq} \int_{\vk} \frac{\nb(\e_{\vp} -\mu') 
\nb(\e_{\vk} -\mu') \nb(\e_{\vq} -\mu')}
{\e_{\vp} +\e_{\vk} -\e_{\vq} -\e_{\vp +\vk -\vq}}  \, \, .
\end{multline}
\end{widetext}
The expression for $\O'$ obtained in Ref \cite{kanno3} is essentially
the same as eq \eqref{kanno_O'_prtbn}.  

\emph{Comparison \& remarks}.\\
The parameter $\mu'(\tc)$ corresponds to the infrared cutoff used in
the the non-selfconsistent regulariztion of \cite{our_f02}.  Eq
\eqref{kanno_mu'} gives the same result for the parameter $\mu'(\tc)$
as that obtained in \cite{our_f02} for the chemical potential $\mu$
using $\mu = \S(0,0)$.
The reason for this is that $\del{F}/\del{N_0}$ represents the energy
of a zero-momentum quasiparticle with respect to the chemical
potential, i.e., $\S(0,0)-\mu$.  Therefore eq \eqref{kanno_mu'} is
equivalent to the usual condition $\S(0,0) = \mu$ for the
Bose-condensation point.

The shift $\D\tc$ obtained using eqs \eqref{kanno_N},
\eqref{kanno_mu'}, \eqref{kanno_O'_prtbn} is the same as that obtained
by our non-selfconsistent regularization scheme in \cite{our_f02}.

The distinctive idea in Kanno's work is the introduction of a modified
thermodynamic function $\O'$ such that $-\del\O'/\del\mu'$ replaces
the right side in the criticality condition $N = (V/\l^3)
\zeta(\tfrac{3}{2})$.  This bypasses the issue of self-consistency and
the modification of the spectrum at $\tc$, and the use of the chemical
potential as a cutoff folllows naturally.

\section{Summary, Further Calculations}   \label{sect_further}

To summarise, this article reviews aspects of the calculation of the
$\tc$ shift due to a small repulsive interaction in a Bose gas.  This
question has generated a number of calculations involving diverse
techniques; the present discussion is a first step toward developing
an organized overview of the different approaches.

The issue that needs to be resolved urgently, in my opinion, is the
possibility of a logarithm-like correction to the leading shift in
$\tc$.  Presumably, a better understanding of the spectrum at the
critical point (both at the infrared and the ultraviolet) should help
settle this issue.  Two directions need to pursued in this regard.

First, as suggested by Stoof, an improved RG calculation, taking into
account the momentum dependence of the effective interaction, would
improve the ultraviolet behavior of the spectrum in his formalism.
Also, it might be worth retaining the wavefunction renormalization
(critical index $\eta$), which changes the infrared spectrum at $\tc$
from $\sim{k}^2$ to $\sim{k}^{2-\eta}$.  This is neglected in
\cite{stoof2} on the grounds that $\eta$ is small.

A second approach suggests itself when we note that the quasiparticle
treatments \cite{our_f02,baym_bigpaper,baym_prl} have all calculated
self-energies in terms of the two-body $t$-matrix $t = 4\pi{a}/m$.
This quantity describes all possible collisions between \emph{two}
particles, without taking into account the fact that the surrounding
medium has an effect on these collisions.  Many-body corrections
arising from the surrounding gas can be treated by using the many-body
$T$-matrix, which is the sum of ladder diagrams.  The two-body
$t$-matrix has no temperature-dependence, while the
temperature-dependence of the effective potential (vanishing at $\tc$)
is captured by the many-body $T$-matrix
\cite{stoof_variational,stoof2,shi+griffin}.  Stoof suggests
\cite{stoof_private} that this might explain why the quasiparticle
analysis (perturbative treatment of spectrum modification)
over-adjusts the infrared spectrum.

Determining the self-energies and quasiparticle spectra in terms of
the many-body $T$-matrix has been attempted \cite{our_unpubl}, using
the relations provided by Shi \& Griffin \cite{shi+griffin} between
the bare potential $U$, the $t$-matrix, and the many-body $T$-matrix.
Our attempt was unsuccessful due to the difficulty of disentangling
multiple frequency and momentum inter-dependences at second-order.

Finally, I mention some other work that would be interesting to the
$\tc$ community.
\begin{itemize}
\item One possibility is to use finite-size scaling ideas, in
conjunction with second-order perturbation theory in the
grand-canonical ensemble.  This would be a complementary to the
calculation of Mueller \& Baym \cite{mueller_baym_finite}, which uses
finite-size scaling with first-order perturbation theory in the
canonical ensemble.
\item It would be nice to see Schakel's calculation of the
thermodynamic potential, just below $\tc$, done with the spectrum
treated more properly, either with a Bogoliubov-like spectrum or with
the $E_\vk(\tc) \sim k^{2-\eta}$ spectrum.
\item As mentioned in section \ref{sect_background}, a detailed
comparison is required between the $\d$-expansion work \cite{delta}
and other approaches to $\tc$.
\end{itemize}

\acknowledgments

Informative discussions with H.T.C. Stoof and F. Lalo\"e were made
possible by the generosity of ECT*, Trento (Italy).  Adriaan
M.J. Schakel answered many e-mail questions.  My PhD advisor,
A. E. Ruckenstein, introduced me to the problem, and the calculations
in \cite{our_f02} were performed under his guidance.


\end{document}